%% file: main.tex
\documentclass[fleqn, 10pt]{wlscirep}
\usepackage[utf8]{inputenc}
\usepackage[T1]{fontenc}
\usepackage{graphicx}
\usepackage{subcaption}
\usepackage[edges]{forest}
\usepackage{xcolor}
\usepackage{fontawesome5}
\usepackage{placeins}
\title{A multi-scenario EEG dataset for auditory attention decoding in naturalistic multi-talker environments}

\author[1]{Shu Peng}
\author[1]{Rui Liu}
\author[1]{Yufei Zhang}
\author[1]{Wenlong You}
\author[1]{Zhige Chen}
\author[1]{Jiachen Xi}
\author[1]{Qiyuan Sun}
\author[2]{Yan Liu}
\author[1]{Kay Chen Tan}
\author[1,*]{Jibin Wu}\affil[1]{Department of Data Science and Artificial Intelligence, The Hong Kong Polytechnic University, Hong Kong SAR, China}
\affil[2]{Department of Computing, The Hong Kong Polytechnic University, Hong Kong SAR, China}

\affil[*]{jibin.wu@polyu.edu.hk}

\begin{abstract}
Understanding how the brain selectively follows relevant speech amid competing voices is a central challenge in auditory neuroscience and a key step toward neuro-steered hearing technologies. However, most open-source Electroencephalography (EEG) datasets for Auditory Attention Decoding (AAD) use idealized single-competing-talker paradigms that oversimplify the acoustic, spatial, and semantic structure of everyday communication. To capture this ecological complexity, we introduce the SoundBubble-EEG dataset: a high-density 128-channel EEG resource comprising more than 25 hours of recordings from 30 participants. The paradigm requires listeners to selectively attend to a dynamic target speaker group, a designated ``sound bubble'', amid competing multi-speaker distractor bubbles across three realistic scenarios: a restaurant, a home TV viewing, and a meeting discussion. By bridging the gap between constrained laboratory protocols and real-world auditory scenes, this dataset enables investigations of multi-talker speech comprehension, neural speech tracking, and cross-scenario generalization. It also provides a benchmark for AAD algorithms under realistic acoustic and semantic variability and may support auditory neuroscience and the development of neuro-steered hearing technologies.
\end{abstract}
\begin{document}
    \flushbottom
    \maketitle

    \thispagestyle{empty}

    \section*{Background}

    The human auditory system possesses a remarkable ability to isolate a specific sound source from a complex, noisy background, a computational feat famously known as the "cocktail party effect"~\cite{cherry1953some, bronkhorst2015cocktail}. Understanding how the brain achieves this selective attention is a central challenge not only in cognitive neuroscience but also in engineering and artificial intelligence. Deciphering how we track attended speech helps researchers probe the fundamental principles of auditory scene analysis and cognitive resource allocation~\cite{mesgarani2012selective,o2015attentional}. Furthermore, this knowledge is the foundation of AAD, an engineering frontier aimed at inferring a listener's attentional focus directly from neural activity. EEG is uniquely suited for this research because its high temporal resolution captures the rapid, millisecond-level dynamics of natural speech comprehension, bridging the gap between basic neuroscience and translational applications like neuro-steered hearing aids~\cite{biesmans2016auditory, kaongoen2023future, geirnaert2021electroencephalography}.

    However, our understanding of real-world multi-talker speech comprehension and the development of robust AAD technologies both remain heavily constrained by the limitations of existing open-access datasets. Most public EEG datasets relevant to auditory attention rely on highly idealized experimental paradigms. They typically employ a basic "competing speaker" setup where a single target talker is masked by a single interfering talker (e.g., one distinct voice on the left versus one on the right)~\cite{fuglsang_2018_1199011, das_2019_4004271, deng2019impoverished, holtze2022ear, li2022esaa}. While these datasets are valuable for establishing baseline algorithms, these artificial conditions do not reflect the acoustic complexity of daily life. In real-world social environments, acoustic interference rarely comes from a single, relentless competing voice. Instead, listeners frequently encounter "Sound Bubbles" - clusters of multiple individuals engaging in a shared conversation~\cite{chen2024hearable}. The cognitive load and neural dynamics required to filter out a collective group conversation differ significantly from those used to suppress a single steady voice, yet this realistic scenario remains largely unexplored in available datasets.
    
    To bridge the gap between idealized laboratory protocols and ecological reality, we present a high-density EEG dataset designed around a novel "Sound Bubble" paradigm. Specifically, we replace the traditional single-interfering-talker model with acoustic environments containing multiple speakers discussing a coherent theme. This design requires participants to direct their attention to the semantic content of a target group while suppressing a realistic, multi-speaker distractor group, thereby maximizing ecological validity. The dataset captures EEG responses from 30 participants across three carefully designed scenarios that mimic common daily environments: (1) a Meeting scenario with moderate spatial separation in a low-noise background; (2) a Restaurant scenario characterized by high ambient noise and complex background dynamics; and (3) a living room setting, where a fixed TV source competes with side conversations. 
    
    By providing over 25 hours of high-quality EEG recordings alongside precise behavioral logs and fully spatialized audio stimuli, this dataset delivers a multifaceted resource with broad interdisciplinary impact. For cognitive neuroscientists, it offers a rich, naturalistic foundation to investigate neural speech tracking, dynamic cognitive load, and listening effort under ecologically valid conditions. For machine learning researchers and biomedical engineers, it serves as a rigorous, real-world benchmark to evaluate the robustness and cross-scenario generalization of AAD algorithms. Ultimately, by capturing the true acoustic complexity of everyday social interactions, this dataset stands to accelerate both our fundamental understanding of auditory cognition and the translational development of next-generation brain-computer interfaces, such as neuro-steered hearing aids and intelligent assistive listening technologies.

    \begin{figure}[!htbp]
        \centering
        \includegraphics[width=\linewidth]{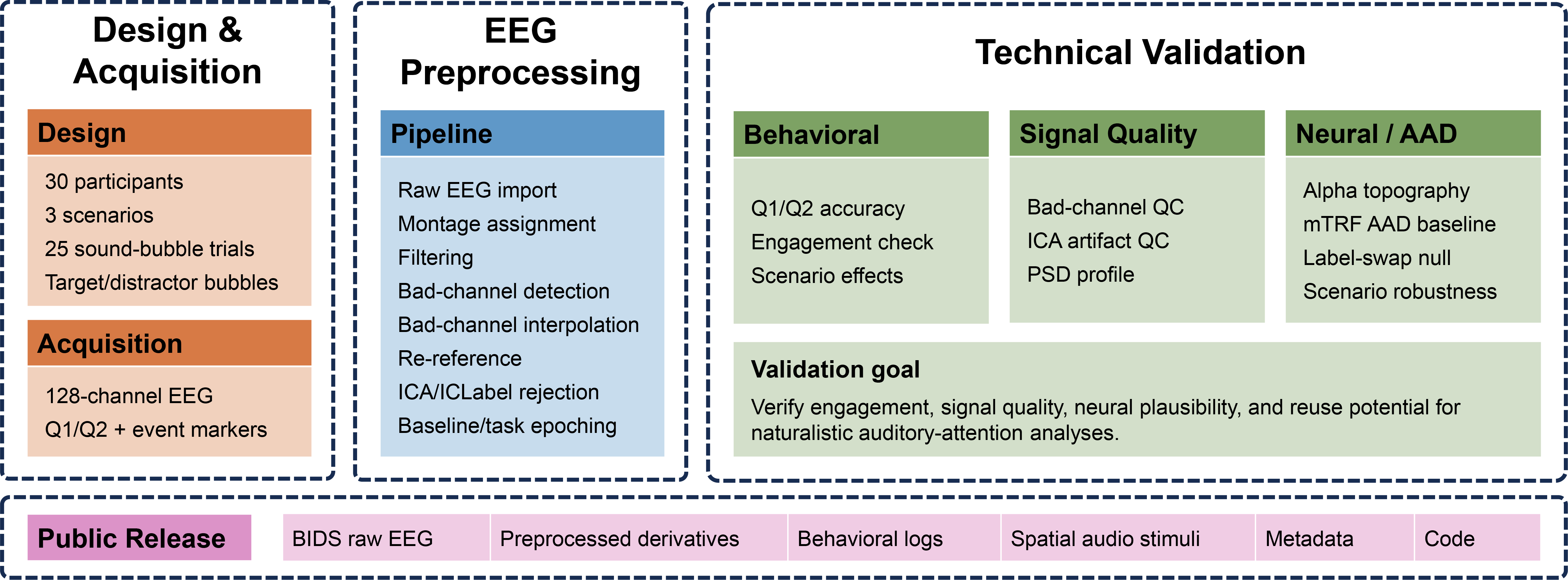}
        \caption{Overview of the SoundBubble-EEG dataset generation, validation, and public release workflow. Participants are engaged in three ecologically valid listening scenarios: meeting, restaurant, and living room (orange). The acquired EEG signals were subsequently processed using a standardized preprocessing pipeline (blue) and underwent rigorous validation, including behavioral assessments, signal quality evaluations, and neural/application-level analyses (green). Finally, the dataset was publicly released in a BIDS-compliant format, accompanied by comprehensive analysis code to facilitate reproducibility and further research (purple).} 
        \label{fig:scheme}
    \end{figure}

    \begin{figure}[!htbp]
        \centering
        \begin{subfigure}{\linewidth}
            \includegraphics[width=\linewidth]{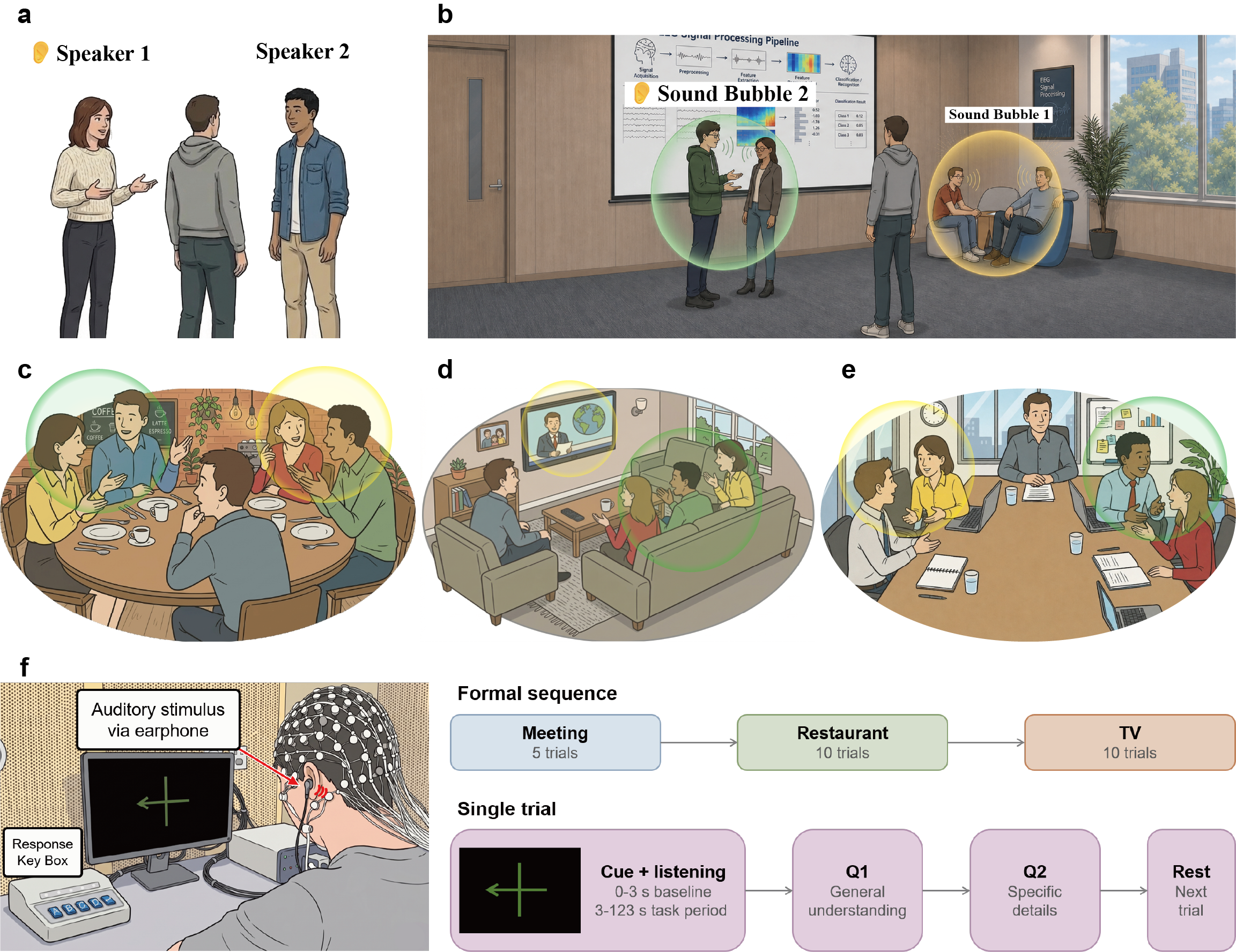}
            \phantomsubcaption \label{fig:soundbubble_a}
            \phantomsubcaption \label{fig:soundbubble_b}
            \phantomsubcaption \label{fig:soundbubble_c}
            \phantomsubcaption \label{fig:soundbubble_d}
            \phantomsubcaption \label{fig:soundbubble_e}
            \phantomsubcaption \label{fig:soundbubble_f}
        \end{subfigure}
        \caption{Illustration of the "Sound Bubble" paradigm and the experimental procedure. 
        (a) The traditional competing-talker paradigm, which typically features two isolated single speakers (e.g., one on the left versus one on the right) without any environmental background sound. The \includegraphics[height=1em]{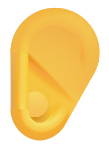} icon indicates the target sound source that the participant is instructed to focus on. (b) The proposed "Sound Bubble" paradigm replaces individual talker scenarios with clusters of multiple speakers, referred to as "Sound Bubbles", each engaged in discussions centered on a coherent theme and positioned within a realistic acoustic environment. This design requires the listener to direct attention to a target bubble while suppressing a multi-speaker distractor bubble and ambient noise, significantly enhancing ecological validity.
        (c)--(e) Three ecologically valid listening scenarios were designed for the dataset, in which the green and yellow bubbles indicate the target (attended) and non-target (unattended) sound bubbles, respectively: (c) Restaurant scene characterized by high ambient noise and two competing conversational sound bubbles; (d) Home TV viewing scene, where a fixed media source (TV) competes with side family conversations in a low-noise background; and (e) Meeting scene, where multiple participants form two discussion groups and each speak from a different direction. (f) Experimental procedure. The formal session followed the actual scenario order of Meeting (5 trials), Restaurant (10 trials), and TV (10 trials). In each trial, a black screen with a green directional cue indicated the target sound source and remained visible throughout listening; the first 3 seconds served as baseline and the following 120 seconds as the active-listening task period. Participants then answered a general-comprehension question (Q1) and a specific-details question (Q2), followed by a self-paced rest before the next trial.}
    \end{figure}

    \section*{Methods}
    The overall data generation and validation pipeline for the SoundBubble-EEG dataset is illustrated in Figure~\ref{fig:scheme}. The workflow is organized around three key components: experimental design and data acquisition, offline EEG preprocessing, and data quality validation. The final public release includes raw EEG signals that follow the Brain Imaging Data Structure (BIDS), preprocessed derivatives, behavioral logs, spatial audio stimuli, metadata, and analysis code~\cite{gorgolewski2016brain}.

    \begin{table}[ht]
        \centering
        \caption{\textbf{Participant demographics and behavioral accuracy.} This
        table details the age, sex, handedness, and formal-trial behavioral
        accuracy for each participant (`M' = Male, `F' = Female; `R' =
        Right, `L' = Left). Q1 assesses general comprehension of the attended stream, while Q2 focuses on specific details. }
        \label{tab:participants}
        \begin{tabular}{|c|c|c|c|c|c|}
            \hline
            \textbf{Subject ID} & \textbf{Age}        & \textbf{Sex}     & \textbf{Handedness} & \textbf{Q1 Mean Acc. (\%)} & \textbf{Q2 Mean Acc. (\%)} \\
            \hline
            sub-01         & 26    & M    & R    & 100.0    & 100.0    \\
            sub-02         & 25    & M    & R    & 96.0     & 76.0     \\
            sub-03         & 26    & M    & R    & 100.0    & 84.0     \\
            sub-04         & 19    & F    & R    & 100.0    & 80.0     \\
            sub-05         & 26    & M    & L    & 96.0     & 60.0     \\
            sub-06         & 24    & M    & R    & 96.0     & 80.0     \\
            sub-07         & 27    & M    & R    & 96.0     & 76.0     \\
            sub-08         & 28    & F    & R    & 100.0    & 76.0     \\
            sub-09         & 25    & M    & R    & 100.0    & 56.0     \\
            sub-10         & 27    & F    & R    & 92.0     & 76.0     \\
            sub-11         & 27    & M    & R    & 96.0     & 84.0     \\
            sub-12         & 24    & M    & R    & 100.0    & 84.0     \\
            sub-13         & 30    & M    & R    & 96.0     & 84.0     \\
            sub-14         & 18    & M    & R    & 100.0    & 88.0     \\
            sub-15         & 28    & M    & R    & 96.0     & 80.0     \\
            sub-16         & 27    & M    & R    & 92.0     & 80.0     \\
            sub-17         & 25    & M    & R    & 96.0     & 80.0     \\
            sub-18         & 25    & M    & R    & 92.0     & 52.0     \\
            sub-19         & 25    & M    & R    & 100.0    & 84.0     \\
            sub-20         & 26    & M    & R    & 100.0    & 84.0     \\
            sub-21         & 28    & M    & R    & 96.0     & 64.0     \\
            sub-22         & 27    & M    & R    & 100.0    & 60.0     \\
            sub-23         & 22    & F    & R    & 100.0    & 68.0     \\
            sub-24         & 28    & F    & L    & 96.0     & 64.0     \\
            sub-25         & 22    & F    & R    & 96.0     & 84.0     \\
            sub-26         & 27    & F    & R    & 100.0    & 84.0     \\
            sub-27         & 26    & F    & R    & 92.0     & 76.0     \\
            sub-28         & 24    & F    & R    & 96.0     & 88.0     \\
            sub-29         & 29    & F    & R    & 96.0     & 72.0     \\
            sub-30         & 27    & M    & R    & 100.0    & 64.0     \\
            \hline
            \textbf{Mean $\pm$ Standard Deviation}  & \textbf{25.6 $\pm$ 2.66} & \textbf{10F/20M} & \textbf{28R/2L}     & \textbf{97.2 $\pm$ 2.81}        & \textbf{76.3 $\pm$ 10.91}        \\
            \hline
        \end{tabular}
    \end{table}
    
    \subsection*{Participants}
    A total of 30 healthy volunteers (mean age = 25.6 $\pm$ 2.66 years; 10 females), comprising students and staff from The Hong Kong Polytechnic University, were recruited for this auditory attention EEG experiment. All participants were native Chinese speakers. The cohort consisted of 28 right-handed and 2 left-handed individuals. All participants self-reported normal hearing and completed practice trials to familiarize themselves with the spatialized audio cues. None reported a history of neurological or psychiatric disorders. The study was approved by the Institutional Review Board of The Hong Kong Polytechnic University (Reference No. HSEARS20230628001). Written informed consent was obtained from all participants prior to the experiment, and they were compensated for their participation.
    
    \subsection*{Stimuli}
    The experiments are designed based on the "Sound Bubble" paradigm that simulate ecologically valid, multi-speaker acoustic environments. As illustrated in Figure~\ref{fig:soundbubble_a} and \ref{fig:soundbubble_b}, unlike traditional dichotic listening tasks that employ single competing speakers, a "Sound Bubble" is defined as a cohesive group of two or more speakers discussing a specific theme. A scenario featuring two sound bubbles and an additional spatial interference source creates a cocktail-party-like acoustic environment. This design aligns closely with realistic listening situations, eliciting richer neural responses for the study of auditory attention and speech tracking. We generated 25 spatial audio segments following the "Sound Bubble" paradigm, categorized into three distinct auditory scenarios that vary in environmental noise and spatial configuration:

    \begin{table}[ht]
        \centering
        \caption{\textbf{Spatial configurations of the 25 auditory stimulus
        templates.} Azimuth angles are expressed relative to the
        listener-facing direction ($0^{\circ}$ = front; negative values = left;
        positive values = right). Elevation is $0^{\circ}$ (eye level) for all
        sources.}
        \label{tab:spatial_config} \resizebox{\textwidth}{!}{
        \begin{tabular}{|c|c|c|c|c|c|}
            \hline
            \textbf{Stimulus ID} & \textbf{Scenario} & \textbf{Listener Facing} & \textbf{Sound Bubble 1 (Pos)} & \textbf{Sound Bubble 2 (Pos)} & \textbf{Noise Source (Pos)} \\
            \hline
            S06015 & Meeting & $0^\circ$ & $-85^\circ$ & $+27^\circ$ & $+57^\circ$ \\
            S06746 & Meeting & $0^\circ$ & $-53^\circ$ & $+79^\circ$ & $-57^\circ$ \\
            S06943 & Meeting & $0^\circ$ & $+77^\circ$ & $-36^\circ$ & $-61^\circ$ \\
            S08364 & Meeting & $0^\circ$ & $+72^\circ$ & $-67^\circ$ & $-50^\circ$ \\
            S08440 & Meeting & $0^\circ$ & $+61^\circ$ & $-56^\circ$ & $-17^\circ$ \\
            S00358 & Restaurant & $0^\circ$ & $-36^\circ$ & $+77^\circ$ & $+44^\circ$ \\
            S01111 & Restaurant & $0^\circ$ & $+83^\circ$ & $-53^\circ$ & $+68^\circ$ \\
            S02021 & Restaurant & $0^\circ$ & $-40^\circ$ & $+72^\circ$ & $+33^\circ$ \\
            S02853 & Restaurant & $0^\circ$ & $+66^\circ$ & $-57^\circ$ & $-42^\circ$ \\
            S03085 & Restaurant & $0^\circ$ & $-29^\circ$ & $+75^\circ$ & $-32^\circ$ \\
            S03550 & Restaurant & $0^\circ$ & $+41^\circ$ & $-75^\circ$ & $+5^\circ$ \\
            S04031 & Restaurant & $0^\circ$ & $-79^\circ$ & $+82^\circ$ & $+48^\circ$ \\
            S04635 & Restaurant & $0^\circ$ & $-29^\circ$ & $+73^\circ$ & $+131^\circ$ \\
            S05016 & Restaurant & $0^\circ$ & $-87^\circ$ & $+50^\circ$ & $-23^\circ$ \\
            S05606 & Restaurant & $0^\circ$ & $+44^\circ$ & $-70^\circ$ & $-73^\circ$ \\
            S00013 & TV & $0^\circ$ & $0^\circ$ & $-89^\circ$ & $-56^\circ$ \\
            S00362 & TV & $0^\circ$ & $+2^\circ$ & $+85^\circ$ & $-7^\circ$ \\
            S01282 & TV & $0^\circ$ & $-9^\circ$ & $+50^\circ$ & $-6^\circ$ \\
            S01563 & TV & $0^\circ$ & $+2^\circ$ & $+80^\circ$ & $+1^\circ$ \\
            S02144 & TV & $0^\circ$ & $+3^\circ$ & $+81^\circ$ & $-4^\circ$ \\
            S02525 & TV & $0^\circ$ & $-4^\circ$ & $+66^\circ$ & $-3^\circ$ \\
            S03502 & TV & $0^\circ$ & $0^\circ$ & $-80^\circ$ & $+3^\circ$ \\
            S04395 & TV & $0^\circ$ & $-8^\circ$ & $+51^\circ$ & $-25^\circ$ \\
            S04407 & TV & $0^\circ$ & $+2^\circ$ & $+85^\circ$ & $+92^\circ$ \\
            S04421 & TV & $0^\circ$ & $-10^\circ$ & $+59^\circ$ & $+13^\circ$ \\
            \hline
        \end{tabular}%
        }
    \end{table}

    \begin{itemize}
    \item \textbf{Restaurant Scene (10 segments):} Simulates a bustling public dining environment with typical restaurant babble and high ambient noise (e.g., clattering dishes). Two competing sound bubbles were spatially rendered at template-specific lateralized azimuths on opposite sides of the listener.
    \item \textbf{Home TV Viewing Scene (10 segments):} Simulates a living room environment with low background noise. One sound bubble (representing a TV program) was rendered near the frontal direction, while another (simulating a side family conversation) was lateralized according to the scene-specific spatial template.
    \item \textbf{Meeting Discussion Scene (5 segments):} Models a professional meeting with multiple participants speaking from template-specific lateralized directions against a low background noise.
    
    \end{itemize}
    All spatial audio stimuli were synthesized using the CIPIC Head-Related Transfer Function (HRTF) database~\cite{algazi2001cipic} to simulate realistic interaural time differences (ITD) and interaural level differences (ILD). Speech samples were sourced from the AISHELL4 dataset~\cite{fu2021aishell} (sampled at 16 kHz). Scene-appropriate background noise was sourced from the DEMAND (Diverse Environments Multichannel Acoustic Noise Database)~\cite{thiemann2013diverse}, encompassing restaurant, home, and meeting recordings. The signal-to-noise ratio (SNR) was carefully controlled to maintain a challenging cognitive load. Finally, all audio stimuli were presented via SONY MDR-EX15LP in-ear headphones.

    \subsection*{Experimental Procedure}

    The formal experiment consisted of 25 trials, with each trial featuring one "Sound Bubble" spatial audio segment. Participants were seated comfortably in a sound attenuated and electromagnetically shielded room. Prior to the formal experiment, participants completed several practice trials using supplementary audio segments to familiarize themselves with the spatial audio paradigm and the experimental procedure. Each trial was initiated by a participant button press. As illustrated in Figure~\ref{fig:soundbubble_f}, a fixation cross, accompanied by a visual cue arrow indicating the target direction (e.g., Left, Right, or Front), appeared centrally on the screen, followed by the audio playback. The audio segments lasted approximately 126 seconds. For downstream EEG analysis, the initial 3-second no-speech cue/background period served as the baseline, while the subsequent 120-second active-listening window (3--123 seconds post-onset) constituted the task period. Participants were instructed to maintain visual fixation and attend to the conversation originating from the cued direction. Upon completion of the audio, participants answered two multiple-choice questions regarding the attended stream: one assessing general semantic understanding and one targeting specific details. Participants then took a short, self-paced rest break before initiating the next trial. 
    
    To optimize the overall experimental duration and manage participant fatigue, the experiment was organized into three distinct blocks with a deliberately designed trial distribution and a fixed presentation order. Specifically, the Meeting scenario, which shares a similar spatial configuration with the Restaurant setting but features a significantly lower auditory cognitive load due to reduced background noise, was allocated 5 trials to balance the total experiment time. The other two scenarios were allocated 10 trials each. Furthermore, rather than randomizing the conditions, we employed a fixed presentation sequence (Meeting, Restaurant, and Home TV) to strategically allocate participants' cognitive resources. The Meeting block was administered first, serving as an acclimatization phase that allowed participants to build task proficiency in a low-demand environment. The highly challenging Restaurant block was placed in the middle, leveraging participants' peak cognitive energy and operational familiarity. Finally, the relatively less demanding Home TV block was presented last to accommodate potential end-of-experiment cognitive fatigue. Extended breaks were provided between blocks, during which
    electrode impedances were checked and saline solution was replenished if
    necessary to keep impedances below 50 k$\Omega$.

    \subsection*{Data Acquisition}
    The experiment was conducted at The Hong Kong Polytechnic University. The EEG data were collected
    using the Geodesic EEG System (Magstim EGI, Eugene, OR, USA), comprising a Net
    Amps 400 amplifier and a 128-channel HydroCel Geodesic Sensor Net (HCGSN).
    The electrode layout followed the standard HCGSN montage. All channels were referenced
    online to the Vertex (Cz), with the COM sensor (located posterior to Cz)
    serving as the ground. Continuous EEG data were recorded using Net Station software
    with a sampling rate of 250 Hz. The electrode impedance was kept below 50 k$\Omega$ prior to recording. Stimulus presentation and event
    marking were controlled using E-Prime 2.0 software (Psychology Software
    Tools, Pittsburgh, PA). Upon completion of the experiment, raw EEG data were
    exported in .mff format, and participants' behavioral data were logged as .txt files
    for subsequent analysis.

    \subsection*{Data Preprocessing}
    EEG data preprocessing was performed offline using Python (version 3.10.19)
    and the MNE-Python~\cite{gramfort2013meg} library (version 1.11.0).

    \subsubsection*{Preprocessing Pipeline}
    \begin{enumerate}
        \item \textbf{Montage and Channel Inspection:} The standard GSN-HydroCel-128
            montage was applied. The VREF channel was designated as a miscellaneous
            channel.

        \item \textbf{Bad Channel Detection:} Noisy channels were automatically
            identified using the PyPREP (version  0.5.0) pipeline~\cite{bigdely2015prep},
            which implements the RANSAC (Random Sample Consensus) method to
            detect channels with low correlation to their neighbors. Identified bad
            channels were marked for interpolation.

        \item \textbf{Filtering:} A band-pass filter was applied between 0.1 and
            40 Hz using a finite impulse response (FIR) filter design to remove
            slow drifts and high-frequency noise~\cite{widmann2015digital}. The
            local power-line frequency (50 Hz) was specified in the PyPREP
            configuration during bad-channel assessment, and line-frequency
            activity was further excluded from the saved derivatives by the
            40-Hz low-pass cutoff.

        \item \textbf{Interpolation and Re-referencing:} Bad channels were
            interpolated using spherical spline interpolation\cite{perrin1989spherical}. The data were then
            re-referenced to the common average reference.

        \item \textbf{Artifact Removal:} Independent Component Analysis (ICA)
            was performed to remove ocular, muscular, and other non-brain artifacts. A copy of the
            data (high-pass filtered at 1.0 Hz to ensure stable decomposition) was
            used to fit the ICA model. The pipeline first attempted Picard configured in extended Infomax mode~\cite{ablin2018faster,lee1999independent}; if this fit failed, the same model was refit using extended Infomax~\cite{lee1999independent}. The \textit{mne-icalabel} package was used to
            automatically classify independent components. Components classified
            as "eye blink", "eye movement", "muscle", "heart beat", or "channel noise" with a
            probability greater than 0.80 were excluded~\cite{pion2019iclabel}. The obtained unmixing matrix
            was then applied to the 0.1--40 Hz filtered data.

        \item \textbf{Epoching:} Continuous data were segmented into epochs
            relative to the audio onset event. Baseline epochs covered the 3-second
            no-speech cue/background period (0--3 s after audio onset), and task epochs
            covered the 120-second active-listening period (3--123 s after audio onset).
    \end{enumerate}
    The sequential order of the preprocessing pipeline was deliberately designed to prevent the propagation of artifacts across the high-density sensor array. Specifically, bad channel detection and spherical spline interpolation were strictly performed prior to re-referencing. This ensures that extreme noise from broken or loose electrodes is neutralized, preventing it from contaminating the remaining healthy channels during the Common Average Reference (CAR) computation. Furthermore, a dual-filtering strategy was employed for ICA. While the primary dataset was band-pass filtered at 0.1--40 Hz to preserve low-frequency neurophysiological features (e.g., event-related slow waves), the ICA model was fitted exclusively on a 1.0 Hz high-pass filtered copy of the data. This is a well-established best practice, as low-frequency drifts can violate the stationarity assumption required for optimal ICA decomposition. The robust spatial weights derived from the 1.0 Hz data were subsequently applied to the 0.1--40 Hz dataset for artifact rejection.

    \section*{Data Records}
    The complete dataset, encompassing raw EEG recordings, preprocessed derivatives,
    behavioral logs, and stimulus materials, is publicly available at Zenodo
    under the DOI: \href{https://doi.org/10.5281/zenodo.19610627}{10.5281/zenodo.19610627}. All data are organized according to the BIDS
    standards. For the stimulus materials, the spatial audio files representing the three scenario categories used in the experiment are provided within the \texttt{BIDS/stimuli/} directory.

    \input{file_list}

    \begin{figure}[!htbp]
        \centering
        \begin{subfigure}{\linewidth}
            \includegraphics[width=\linewidth]{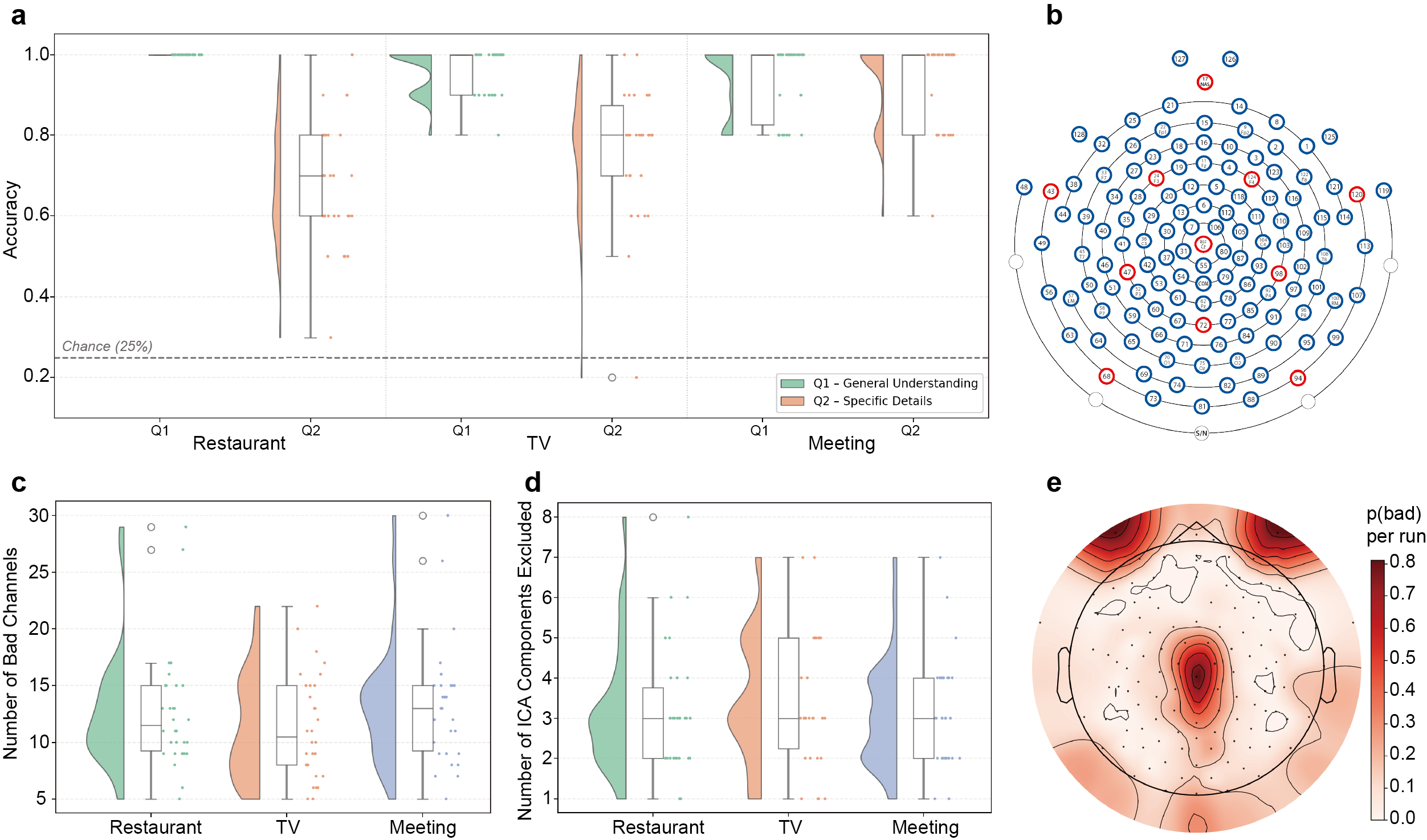}
            \phantomsubcaption \label{fig:technical_validation_a}
            \phantomsubcaption \label{fig:technical_validation_b}
            \phantomsubcaption \label{fig:technical_validation_c}
            \phantomsubcaption \label{fig:technical_validation_d}
            \phantomsubcaption \label{fig:technical_validation_e}
        \end{subfigure}
        \caption{Behavioral performance and data quality assessment. (a) Behavioral accuracy for general understanding (Q1) and specific details (Q2) across the three acoustic conditions (Restaurant, TV, Meeting). The dashed line indicates chance level (0.25).
        (b) Sensor layout of the 128-channel high-density EEG net.
        (c) Distribution of interpolated bad channels per run across three acoustic scenarios.
        (d) Number of artifact independent components excluded via ICA per run across three acoustic scenarios.
        (e) Spatial probability heatmap of bad channels across all recordings.}
    \end{figure}

    \section*{Technical Validation}
    To ensure the reliability and neurophysiological validity of the dataset, we
    performed a comprehensive quality assessment covering behavioral performance,
    signal quality, event-related oscillatory dynamics, and application-level AAD performance.

    \subsection*{Behavioral Performance}
    To verify participant engagement and the successful manipulation of auditory attention, we analyzed the response accuracy for the comprehension questions presented after each formal trial, excluding the practice trials used for familiarization. As depicted in Figure \ref{fig:technical_validation_a}, participants demonstrated robust performance across all three ecologically valid scenarios (Restaurant, TV, and Meeting), consistently scoring well above the 25\% chance level for Multiple-choice questions. 

    Notably, a distinct performance pattern emerged between the two question types. Accuracy for Q1 (assessing general understanding) reached near-ceiling levels across all conditions, indicating that participants could successfully lock onto and track the primary semantic thread of the target "Sound Bubble" despite realistic distractors. Accuracy for Q2 (assessing specific details) exhibited greater variance and slightly lower median scores, which accurately reflects the increased cognitive demand required to extract fine-grained information in complex, multi-speaker environments. Crucially, even in the highly demanding Restaurant scenario, characterized by dense ambient noise and intense competing sound bubbles, the median accuracy remained high. This compelling behavioral evidence confirms that participants were actively segregating the target sound stream from the background noise, thereby validating the efficacy of the paradigm.

    \subsection*{Signal Quality Assessment}
    We first quantified the number of bad channels identified and interpolated during each run of the RANSAC algorithm (implemented in the PyPREP pipeline). As illustrated in Figure~\ref{fig:technical_validation_c}, the median number of bad channels was consistently maintained at approximately 10 to 13 out of 128 channels across all three scenarios. The highly similar probability density distributions across conditions indicate that the physical impedance of the hydrocel geodesic net was not systematically affected by the varying acoustic environments. While a few statistical outliers were observed (e.g., runs with up to 30 bad channels, primarily due to individual differences in hair thickness or localized cap tension), these were successfully recovered using spherical spline interpolation.

    Furthermore, we mapped the spatial probability of channel degradation across all subjects and runs (Figure~\ref{fig:technical_validation_e}). The heatmap demonstrates a predictable physical pattern typical of saline-based high-density nets, with elevated rejection probabilities at several peripheral or facial electrodes and around midline/vertex sensors where hair interference, cap tension, or saline drying can be more prominent. Some recurrently interpolated channels were not confined exclusively to the outermost ring, so this map should be interpreted as a transparent spatial quality-control profile rather than as evidence for condition-specific degradation. These channels were interpolated before re-referencing, reducing their influence on the cleaned derivatives.

    To verify that the different environmental noise conditions did not introduce confounding physiological artifacts (e.g., increased eye blinks or muscle tension due to louder or more complex backgrounds), we analyzed the number of independent components removed per run. Using the Picard-first ICA procedure with extended Infomax fallback combined with ICLabel, artifactual components with over 80\% classification probability were automatically excluded. The results (Figure~\ref{fig:technical_validation_d}) show a highly stable median of 3 to 4 rejected components per run across three conditions. The absence of significant condition-specific shifts in the ICA rejection distributions confirms that the subjects maintained a consistent behavioral and physiological state throughout the experiment, ensuring that downstream analyses are driven by genuine neural responses rather than systematic artifactual variance.

\begin{figure}[!htbp]
        \centering
        \begin{subfigure}{\linewidth}
            \includegraphics[width=\linewidth]{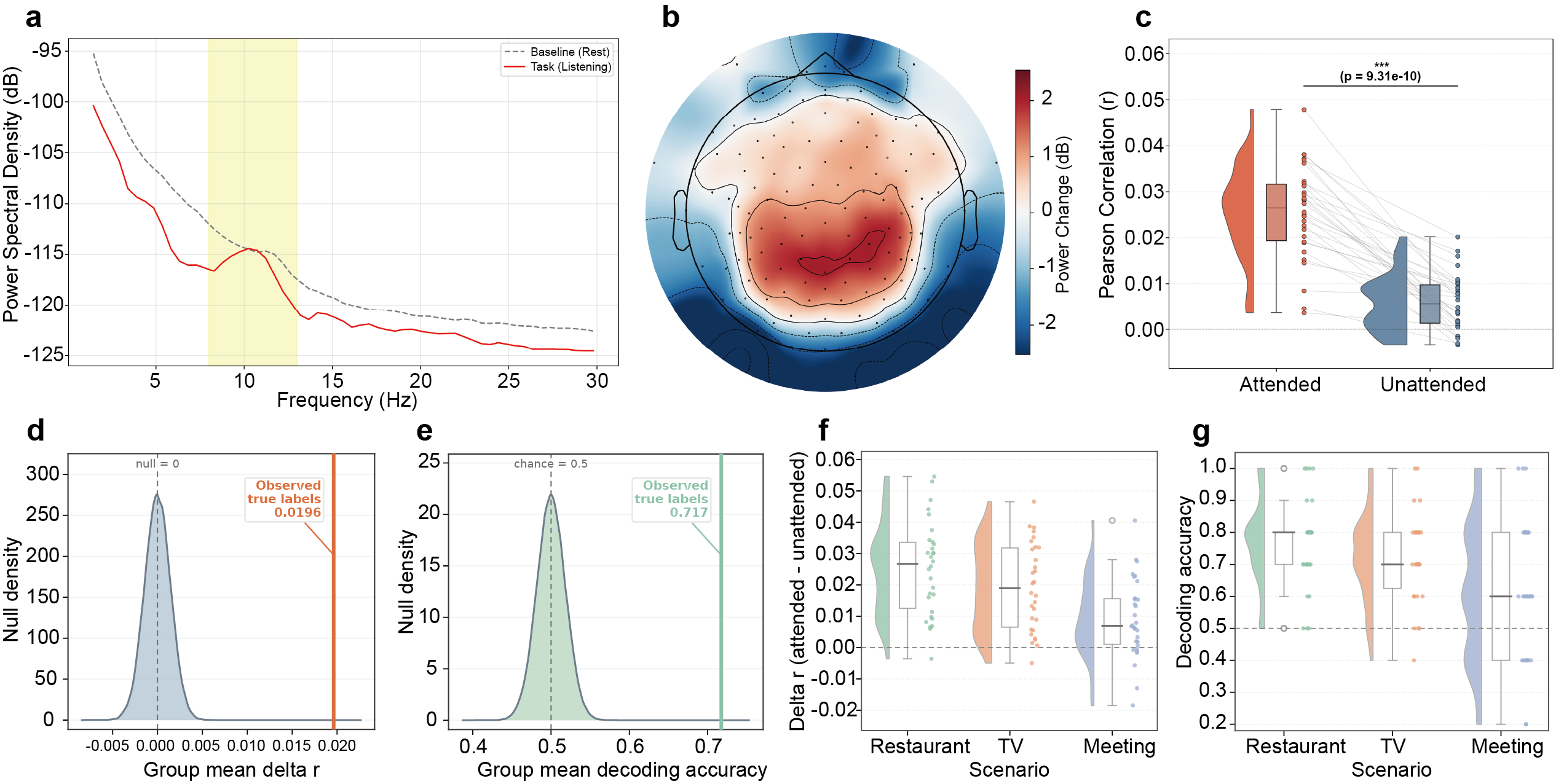}
            \phantomsubcaption \label{fig:technical_validation2_a}
            \phantomsubcaption \label{fig:technical_validation2_b}
            \phantomsubcaption \label{fig:technical_validation2_c}
            \phantomsubcaption \label{fig:technical_validation2_d}
            \phantomsubcaption \label{fig:technical_validation2_e}
            \phantomsubcaption \label{fig:technical_validation2_f}
            \phantomsubcaption \label{fig:technical_validation2_g}
        \end{subfigure}
        \caption{Neurophysiological sanity checks and application-level AAD validation. (a) Grand average global Power Spectral Density (PSD) for Baseline (dashed grey) and Task (solid red) states, with the alpha band (8--13 Hz) highlighted in yellow.
        (b) Topography of normalized alpha power change ($10\log_{10}(\mathrm{Task}/\mathrm{Baseline})$). Blue regions indicate relative alpha suppression, while red regions denote relative alpha enhancement.
        (c) Auditory attention decoding performance using a backward mTRF model. Paired raincloud plots show Pearson correlations ($r$) between EEG-reconstructed and actual speech envelopes, demonstrating significantly higher cortical tracking for attended speech (one-sided Wilcoxon signed-rank test, *** $p = 9.31 \times 10^{-10}$).
        (d,e) Trial-label swap null controls for the group-mean decoding advantage $\Delta r = r_{\mathrm{attend}} - r_{\mathrm{unattend}}$ and single-trial decoding accuracy. Shaded distributions show label-swapped null distributions; vertical colored lines show the observed values computed from the true attended/unattended labels.
        (f,g) Scenario-wise AAD robustness for $\Delta r$ and decoding accuracy across Restaurant, TV, and Meeting scenarios. Scenario-wise panels show subject-level scenario means; Restaurant and TV contained 10 trials per subject, whereas Meeting contained 5 trials per subject.}
    \end{figure}
    \FloatBarrier
    
    \subsection*{Neurophysiological Sanity Checks}
    To confirm that the preprocessed EEG dataset captures genuine neural dynamics, we evaluated the data against well-established frequency-domain and time-domain neurophysiological signatures associated with auditory spatial attention. The global Power Spectral Density (PSD) was computed to compare the active listening state (Task) against the pre-speech cue/background baseline (Baseline). As illustrated in Figure~\ref{fig:technical_validation2_a}, both spectra exhibited a canonical logarithmic 1/f decay across frequencies, confirming the physiological validity of the recordings. The task spectrum was generally lower than the baseline spectrum across broad frequency ranges, consistent with task-related desynchronization and the fact that the baseline was a short cue/background period rather than a resting-state recording. To visualize the spatial distribution of alpha-band modulation, we calculated the normalized Alpha band (8--13 Hz) power change as $10\log_{10}(\mathrm{Task}/\mathrm{Baseline})$ for each channel (Figure~\ref{fig:technical_validation2_b}). The resulting topographic map shows spatially heterogeneous alpha modulation, including alpha suppression over temporal/frontal regions and relative alpha enhancement over parieto-occipital sensors. This posterior alpha increase is consistent with cross-modal visual inhibition during sustained auditory attention in a complex acoustic environment~\cite{jensen2010shaping, foxe2011role}.
    
    \subsection*{Application Baseline}
    To evaluate the practical utility of the dataset and verify that the recorded EEG signals robustly capture the neural mechanisms of selective attention, we performed an AAD analysis. This serves as a standard technical baseline to confirm that the data can be effectively utilized by brain-computer interface (BCI) algorithms to reconstruct the attended speech envelope from single-trial EEG. We implemented a classic backward multivariate Temporal Response Function (mTRF) using Ridge regression~\cite{o2015attentional, biesmans2016auditory}. For each subject, a leave-one-out cross-validation was performed across all 25 independent trials (120 seconds each). By pooling training trials across the Restaurant, TV, and Meeting conditions, we trained a scenario-pooled, trial-independent baseline model and evaluated it on the held-out trial. The model was trained to linearly map the multi-channel EEG signals (time-shifted between -400 ms and 100 ms relative to the audio) to the speech envelope of the attended stream. The decoding performance on the held-out test trial was quantified by calculating the Pearson correlation coefficient ($r$) between the reconstructed envelope and the actual envelopes of both the attended ($r_{\mathrm{attend}}$) and unattended ($r_{\mathrm{unattend}}$) speech streams. From these trial-level correlations, we further derived two complementary AAD metrics for the following validation analyses: the reconstruction advantage $\Delta r = r_{\mathrm{attend}} - r_{\mathrm{unattend}}$ and single-trial decoding accuracy, defined as the proportion of trials for which $r_{\mathrm{attend}} > r_{\mathrm{unattend}}$.

    As shown in Figure~\ref{fig:technical_validation2_c}, at the group level, the reconstructed speech envelopes showed a significantly higher correlation with the attended audio streams compared to the unattended streams. A one-sided Wilcoxon signed-rank test confirmed this attended-stream advantage ($p = 9.31 \times 10^{-10}$), demonstrating a consistent, population-wide neural tracking effect. The absolute single-trial reconstruction correlations were modest (mean $r_{\mathrm{attend}} \approx 0.026$). However, such values are typical in EEG speech-envelope reconstruction and Pearson correlation coefficient (PCC) analyses~\cite{liang2026decoding}, reflecting the inherently low signal-to-noise ratio characteristic of scalp EEG-based AAD estimates~\cite{geirnaert2021electroencephalography}.

    To verify that this AAD effect was specifically tied to the correct attended/unattended labels, we performed a trial-label swap null control on the per-trial mTRF outputs. For each permutation, attended and unattended labels were randomly swapped at the trial level, and the group-level $\Delta r$ and decoding accuracy were recomputed. The observed true-label statistics ($\Delta r = 0.0196$; decoding accuracy = 0.717) lay far outside the corresponding label-swapped null distributions (Figure~\ref{fig:technical_validation2_d},\subref{fig:technical_validation2_e}). None of the 50,000 label-swapped permutations reached the observed values for either metric, yielding permutation $p = 2.0 \times 10^{-5}$ after add-one correction~\cite{phipson2016permutation}. This control confirms that the decoding effect cannot be explained by nonspecific reconstruction bias or chance alignment between EEG and acoustic envelopes.

    Finally, we assessed whether the AAD effect was robust across the three ecological acoustic scenarios rather than being driven by a single condition. Subject-level scenario means were computed separately for Restaurant, TV, and Meeting using the same held-out trial predictions. The reconstruction advantage remained positive in all three scenarios, with mean $\Delta r$ values of 0.0252, 0.0193, and 0.0092 for Restaurant, TV, and Meeting, respectively (Figure~\ref{fig:technical_validation2_f}). Single-trial decoding accuracy showed the same pattern, reaching 0.767, 0.720, and 0.613 across the three scenarios (Figure~\ref{fig:technical_validation2_g}). Although the Meeting condition contained fewer trials per subject (5 trials) than Restaurant and TV (10 trials each), both $\Delta r$ and decoding accuracy remained above their respective null/chance levels. Together, these analyses demonstrate that the dataset captures reliable attended-speech tracking and provides a viable benchmark for AAD algorithms across realistic multi-scenario sound environments.

    \section*{Usage Notes}
    The SoundBubble-EEG dataset is organized according to the BIDS to facilitate ease of use, automated processing, and cross-lab reproducibility. We provide both the raw EEG recordings (in \texttt{.mff} format) and preprocessed derivatives (in \texttt{.fif} format, compatible with MNE-Python) to accommodate diverse analytical needs. Researchers interested in validating novel preprocessing pipelines or artifact rejection algorithms can utilize the raw data, where the original channel signals remain intact for custom quality control. Conversely, those focusing primarily on decoding algorithms or cognitive modeling may prefer the preprocessed versions, which have already undergone rigorous RANSAC-based bad channel interpolation and ICA-based artifact removal. For analysis, we strongly recommend using the MNE-Python library (version 1.11.0 or later), as the provided derivatives and event dictionaries were strictly standardized within this framework. 

    Beyond standard AAD baselines, the rich ecological variations inherent in this dataset present several advanced analytical opportunities. First, the inclusion of three distinct multi-speaker scenarios allows for the rigorous evaluation of \textbf{cross-condition generalization and transfer learning~\cite{lotte2018review}} in BCI algorithms (e.g., training a decoding model on the low-noise Meeting scenario and testing its robustness on the high-noise Restaurant environment). Second, the paired behavioral metrics provide a rare opportunity to investigate the \textbf{brain-behavior relationship}. Researchers can correlate single-trial neural signatures (such as mTRF cortical tracking scores or alpha-band lateralization) with the comprehension accuracy of Q1 (general) versus Q2 (details) to explore the neural correlates of listening effort and cognitive load. Finally, by leveraging the precise source positioning metadata provided in the \texttt{stimuli/metadata/} JSON files, future studies can move beyond conventional binary classification (left vs. right) to explore \textbf{continuous spatial attention decoding} or investigate dynamic functional connectivity networks under varying acoustic complexities.

    \section*{Code Availability}
    The SoundBubble-EEG dataset is available at Zenodo: \href{https://doi.org/10.5281/zenodo.19610627}{https://doi.org/10.5281/zenodo.19610627}. The code for data preprocessing and AAD analyses is available at GitHub: \href{https://github.com/Mashirops/sound-bubble-eeg-code}{https://github.com/Mashirops/sound-bubble-eeg-code}.

    \section*{Acknowledgements}

    This work was partially supported by the Research Grants Council of the Hong Kong SAR (Grant No. C5052-23G, PolyU15215623, PolyU15229824, PolyU25216423, PolyU15217424, and SRFS2526-5S04), the National Natural Science Foundation of China (Grant No. 62306259),  The Hong Kong Polytechnic University (P0056503, P0058445, P0060651).

    \section*{Author contributions}

    S.P., R.L., Y.Z. and J.W. conceived and designed the study. S.P., R.L., W.Y., Z.C., J.X. and Q.S. contributed to data collection and dataset organization. S.P. and R.L. performed data preprocessing and technical validation. S.P. drafted the manuscript. Y.L., K.T., and J.W. supervised the project and revised the manuscript.

    \section*{Competing interests}
    The authors declare no competing interests. 

\end{document}

%% file: file_list.tex

\definecolor{foldercolor}{RGB}{46,96,160}
\definecolor{jsoncolor}{RGB}{180,120,20}
\definecolor{datacolor}{RGB}{40,140,70}
\definecolor{eegcolor}{RGB}{140,60,170}

\newcommand{\fld}[1]{\textcolor{foldercolor}{\faIcon{folder}}~\textcolor{foldercolor}{\textbf{#1}}}
\newcommand{\fldo}[1]{\textcolor{foldercolor}{\faIcon{folder-open}}~\textcolor{foldercolor}{\textbf{#1}}}
\newcommand{\fjson}[1]{\textcolor{jsoncolor}{\faIcon{code}}~{#1}}
\newcommand{\ftsv}[1]{\textcolor{datacolor}{\faIcon{table}}~{#1}}
\newcommand{\feeg}[1]{\textcolor{eegcolor}{\faIcon{brain}}~{#1}}
\newcommand{\ftxt}[1]{\textcolor{gray}{\faIcon{file-alt}}~{#1}}
\newcommand{\fwav}[1]{\textcolor{teal}{\faIcon{music}}~{#1}}
\newcommand{\fellip}[1]{\textcolor{gray}{\faIcon{ellipsis-h}}~\textit{\textcolor{gray}{#1}}}

\begin{figure}[!htbp]
\centering
\begin{forest}
  for tree={
    folder,
    font=\ttfamily\small,
    grow'=0,
    edge={draw, thin},
    l sep=12pt,
    s sep=2pt,
    inner ysep=1pt,
    inner xsep=2pt,
  },
  [{\fldo{Sound\_Bubble/BIDS/}}
    [{\fjson{dataset\_description.json}}]
    [{\ftsv{participants.tsv}}]
    [{\ftxt{README}}]
    [{\ftxt{CHANGES}}]
    [{\fld{sub-01/}}
      [{\fld{eeg/}}
        [{\feeg{sub-01\_task-auditory\_run-01\_eeg.mff}}]
        [{\fjson{sub-01\_task-auditory\_run-01\_eeg.json}}]
        [{\ftsv{sub-01\_task-auditory\_run-01\_channels.tsv}}]
        [{\ftsv{sub-01\_task-auditory\_run-01\_events.tsv}}]
        [{\ftsv{sub-01\_task-auditory\_run-01\_electrodes.tsv}}]
        [{\fjson{sub-01\_task-auditory\_run-01\_coordsystem.json}}]
        [{\fellip{run-02/, run-03/ follow the same pattern}}]
      ]
    ]
    [{\fellip{sub-02/ \ldots{} sub-XX/}}]
    [{\fld{derivatives/}}
      [{\fld{preprocessed/}}
        [{\fjson{dataset\_description.json}}]
        [{\fld{sub-01/}}
          [{\fld{eeg/}}
            [{\feeg{sub-01\_task-auditory\_desc-clean\_epo.fif}}]
            [{\feeg{sub-01\_task-baseline\_desc-clean\_epo.fif}}]
          ]
        ]
        [{\fellip{sub-02/ \ldots{} sub-XX/}}]
      ]
    ]
    [{\fld{sourcedata/}}
      [{\fld{sub-01/}}
        [{\fld{beh/}}
          [{\ftxt{sub-01\_task-auditory\_run-01\_beh.txt}}]
          [{\fellip{run-02/, run-03/ follow the same pattern}}]
        ]
      ]
      [{\fellip{sub-02/ \ldots{} sub-XX/}}]
    ]
    [{\fld{stimuli/}}
      [{\fld{restaurant/}}
        [{\fellip{.wav audio files}}]
      ]
      [{\fld{ts/}}
        [{\fellip{.wav audio files}}]
      ]
      [{\fld{meeting/}}
        [{\fellip{.wav audio files}}]
      ]
    ]
  ]
\end{forest}
\caption{Directory structure of the Sound Bubble dataset organized according to the BIDS specification. Icons indicate file types: \textcolor{foldercolor}{\faIcon{folder}}~directories, \textcolor{eegcolor}{\faIcon{brain}}~EEG data (.mff/.fif), \textcolor{jsoncolor}{\faIcon{code}}~JSON metadata, \textcolor{datacolor}{\faIcon{table}}~tabular data (.tsv) and \textcolor{gray}{\faIcon{file-alt}}~text files. 
Only \texttt{sub-01/run-01} is fully expanded; all other runs and subjects share the same structure.}
\label{fig:file-tree}
\end{figure}